\documentclass{ws-p8-50x6-00}

\input{psfig.sty}

\begin{document}

\title{Dynamical meson-baryon resonances with chiral Lagrangians}

\author{A. Ramos, A. Parre\~no}

\address{Departament ECM, Universitat de Barcelona, Diagonal 647,\\
E-08028 Barcelona, SPAIN}

\author{J.C. Nacher, E. Oset}

\address{Departamento de F\'{\i}sica Te\'orica and IFIC, 
Institutos de Investigaci\'on de
Paterna, Aptdo. Correos 22085,
E-46071 Valencia, SPAIN}

\author{C. Bennhold} 

\address{Center for Nuclear Studies, Department of Physics, \\
The George Washington University, Washington D.C. 20052}       


\maketitle

\abstracts{ 
The s-wave meson-baryon interaction is studied using the
lowest-order chiral Lagrangian in a unitary
coupled-channels Bethe-Salpeter equation.
In the strangeness $S=-1$ sector the 
low-energy $K^- p$ dynamics leads to the dynamical generation
of the $\Lambda(1405)$ as a ${\bar K}N$ state, along with a good description of
the $K^- p$ scattering observables.  At higher energies, 
the $\Lambda(1670)$ is also found to be generated
dynamically  as a $K \Xi$ quasibound state for the first time.
For strangeness $S=0$, it is the $S_{11}(1535)$ resonance
that emerges from the coupled-channels equations,
leading to a satisfactory description of meson-baryon scattering
observables in the energy region around the $S_{11}(1535)$.
We speculate on the possible dynamical generation of $\Xi$ resonances
within the chiral $S=-2$ sector as ${\bar K} \Lambda$ or
${\bar K} \Sigma$ quasibound states.}

\section{Introduction}

Gaining insight into the nature and properties of baryon
resonances is one of the
primary goals of hadron physics. 
In the last years, intensive theoretical and experimental
effort has been devoted to clarify the internal structure of
many resonances and establish whether they behave as 3-quark objects, as
predicted by the constituent quark model, or they can be described mostly
in terms of hadronic degrees of freedom, as meson-baryon quasibound
states.   

In the last decade, chiral perturbation theory (ChPT)
has emerged as a powerful scheme that successfully
explains not only low-energy meson-meson 
dynamics\cite{Ga85,Pi95} but also meson-baryon 
scattering\cite{Eck95,Be95}, provided the interaction is weak as is the case of 
$\pi N$ scattering in the strangeness $S=0$ sector, or $K^+ N$
scattering in the $S=1$ one. However, the validity of
ChPT, which is built as a series expansion in the
meson momentum, is limited to {\it low energies} and, 
in addition, it is not applicable in the vicinity of {\it resonances}, 
which correspond to poles in the T-matrix.
 
Both difficulties can be overcome
within the framework of the chiral Lagrangian, by
combining the information contained in it
with unitarity using non-perturbative techniques. This
has proved to be very successful in both the meson-meson\cite{OOP99,nsd} 
and the meson-baryon sector\cite{Kai95,OR98,OM01,juan01}. 
A review of recent results can be found in Ref. \cite{report}.

\section{Meson-baryon scattering in coupled channels}

Ref.~\cite{Kai95} combined the chiral Lagrangian 
(to lowest and next-to-leading order) and a non-perturbative
resummation by solving a coupled-channels Lippmann-Schwinger equation.
In the case of $K^- p$ scattering, the channels
included were those opened at threshold and, fitting five parameters, the
low-energy scattering observables, as well as the properties of the
$\Lambda(1405)$ resonance, were well reproduced. A similar procedure
was followed in Ref.~\cite{OR98}
in terms of the coupled-channels Bethe-Salpeter equation (BSE)
given by 
\begin{equation} 
T_{ij} = V_{ij} + V_{il} G_{l} T_{lj} \ .
\label{eq:BS} 
\end{equation}
It was shown that the data could be well
reproduced taking the amplitudes
$V_{ij}$ from the lowest order chiral Lagrangian and only one parameter,
the cut-off used to regularize the intermediate
meson-baryon propagator $G_l$. The key difference compared to
Ref.~\cite{Kai95} was the inclusion of
all ten possible meson-baryon states constructed from the octet of
pseudoscalar mesons
and the octet of $1/2^+$ baryons, thus preserving
SU(3) symmetry in the limit of equal baryon and meson
masses. Another advantage of the lowest-order Lagrangian is that the
amplitudes can be factorized
on-shell out of the loop integral, thus reducing the 
coupled-channels problem to a set of coupled algebraic equations.

The unitarization of the chiral amplitudes through the BSE
has been shown to be a particular case of the Inverse Amplitude Method
when the choice of the regulator allows one to include the second-order
chiral contributions by means of only the s-channel
loop\cite{OOP99}. Equivalently, the
BSE is a particular case of the N/D method when the effects of
the unphysical cut (left-hand singularities) are neglected\cite{nsd}.

The amplitudes $V_{ij}$ in Eq.~(\ref{eq:BS}) are easily obtained from
the lowest order meson-baryon interaction Lagrangian 
\begin{equation}
L_1^{(B)} = \langle \bar{B} i \gamma^{\mu} \frac{1}{4 f^2}
[(\Phi \partial_{\mu} \Phi - \partial_{\mu} \Phi \Phi) B
- B (\Phi \partial_{\mu} \Phi - \partial_{\mu} \Phi \Phi)]
\rangle     \ ,
\end{equation}
where $\Phi$ and $B$ are the matrices representing the mesons and baryons,
respectively.
A key ingredient in the BSE is
the loop integral, $G_l$, which in
Ref.~\cite{OR98} reads
\begin{eqnarray}
G_{l} &=& i \, \int \frac{d^4 q}{(2 \pi)^4} \, \frac{M_l}{E_l
(\vec{q}\,)} \,
\frac{1}{k^0 + p^0 - q^0 - E_l (\vec{q}\,) + i \epsilon} \,
\frac{1}{q^2 - m^2_l + i \epsilon} \nonumber \\
&=& \int_{\mid {\vec q} \mid < q_{\rm max}} \, \frac{d^3 q}{(2
\pi)^3} \,
\frac{1}{2 \omega_l (\vec q\,)}
\,
\frac{M_l}{E_l (\vec{q}\,)} \,
\frac{1}{\sqrt{s}- \omega_l (\vec{q}\,) - E_l (\vec{q}\,) + i
\epsilon} \ ,
\label{eq:gprop}
\end{eqnarray}
and has been regularized by means of a
cut-off, $q_{\rm max}$. An alternative approach, followed in
Ref.~\cite{OM01}, is
obtained by making use of dimensional regularization
\begin{eqnarray}
G_{l} &=& i 2 M_l \int \frac{d^4 q}{(2 \pi)^4} \, 
\frac{1}{(P-q)^2 - M_l^2 + i \epsilon} \,
\frac{1}{q^2 - m^2_l + i \epsilon} \nonumber \\
&=& \frac{2 M_l}{16 \pi^2} \left\{ a_l(\mu) +
\ln \frac{M_l^2}{\mu^2} + \frac{m_l^2-M_l^2 + s}{2s} \ln
\frac{m_l^2}{M_l^2} + \right. \nonumber \\
& & \left. \phantom{\frac{2 M}{16 \pi^2}} + \frac{\bar{q}_l}{\sqrt{s}} \ln
\frac{M_l^2
+ m_l^2 - s -2\sqrt{s} \bar{q}_l}{M_l^2 + m_l^2 - s +2\sqrt{s}
\bar{q}_l} \right\} \ ,
\label{eq:gpropdr}
\end{eqnarray}
where $\mu$ is the regularization scale and $\bar{q}_l$ the
on-shell momentum. Obviously, changes in the
cut-off value of Eq.~(\ref{eq:gprop}) can always be accommodated with
changes in the regularization scale $\mu$ and the corresponding 
change in the subtraction constant $a_l(\mu)$ in
Eq.~(\ref{eq:gpropdr}). The dimensional
regularization scheme can extend the model to higher energies,
while the cut-off method limits the range of applicability to energies whose
on-shell momentum ${\bar q}_l$ is smaller than the cut-off value for all
channels.

\section{Strangeness $S=-1$ sector}

\begin{figure}[ht]
\centerline{
\psfig{file=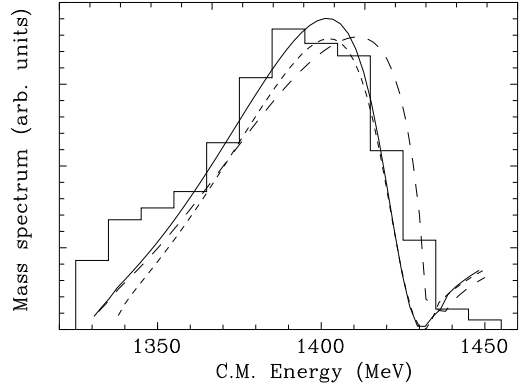,height=5.5cm,angle=0,silent=} 
}
\caption{$\Lambda(1405)$ resonance as obtained from the invariant
$\pi\Sigma$ mass distribution, with the full basis of physical
states
(solid line), omitting the $\eta$ channels (long-dashed line) and
with the isospin-basis (short-dashed line). Experimental
histogram from Ref.~\protect\cite{spec}. Figure taken from
Ref.~\protect\cite{OR98}.
\label{fig:lambda}}
\end{figure}

\begin{figure}
\centerline{
\psfig{file=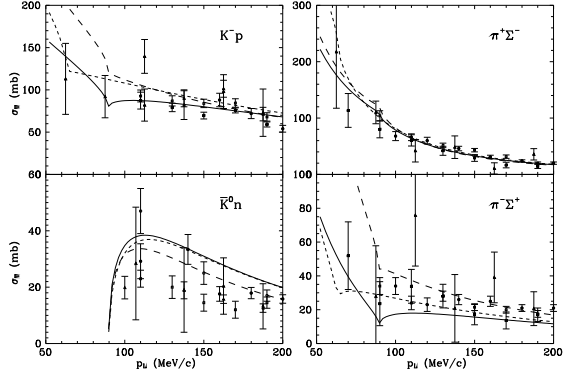,height=5.5cm,angle=0,silent=}
} 
\caption{
$K^-p$ scattering cross sections as functions of the $K^-$
momentum in the lab frame: with the full basis of physical states
(solid line), omitting the $\eta$ channels (long-dashed line) and
with the isospin-basis (short-dashed line). Taken from
Ref.~\protect\cite{OR98}.
\label{fig:kncross}}
\end{figure}

We start this section summarizing the findings of
Ref.~\cite{OR98} for ${\bar K}N$ scattering.
The cut-off method was used
with a value for the decay constant of $f=1.15 f_\pi$ (chosen
between $f_{\pi} = 93$ MeV and $f_{K}=1.22 f_{\pi}$)
and a value $q_{\rm max}=630$ MeV was
obtained by reproducing the threshold branching ratios,
$\gamma=\Gamma (K^- p \rightarrow \pi^+ \Sigma^-)/
\Gamma (K^- p \rightarrow \pi^- \Sigma^+)$, $R_c=
\Gamma (K^- p \rightarrow {\rm charged})/\Gamma (K^- p \rightarrow {\rm all})$
and
$R_n = \Gamma (K^- p \rightarrow \pi^0 \Lambda)/
\Gamma (K^- p \rightarrow {\rm neutral})$.
The above value for $f$ 
gave the best agreement for the
$\Lambda(1405)$ properties seen in the
$\pi \Sigma$ mass spectrum, as shown in
Fig.~\ref{fig:lambda}. In Fig.~\ref{fig:kncross}, we show that
the low-energy scattering cross sections, not used in the fit, are
well reproduced.

In order to assess the range of validity of our approach
we explored the region of higher energies, using the dimensional
regularization scheme with $\mu=630$ MeV. In order to maintain the low-energy
results, we adjust the subtraction constant $a_l$ to
reproduce the value of the loop function $G_l$ at threshold
($\sqrt{s}=m_K+m_N$) calculated with
the cut-off, and we find:
\begin{equation}
\begin{array}{lll}
a_{{\bar K}N}=-1.84~~ & a_{\pi\Sigma}=-2.00~~ &  a_{\pi\Lambda}\,=-1.83 \\
a_{\eta \Lambda}\,\,=-2.25~~ & a_{\eta\Sigma}=-2.38~~ &  a_{K\Xi}=-2.52 \ .
\end{array}
\label{eq:params}
\end{equation}

\begin{figure}
\centerline{
\psfig{file=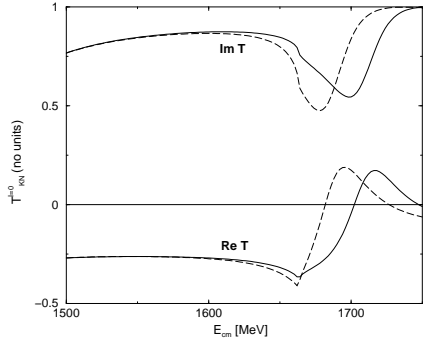,height=5.5cm,angle=270,silent=} 
}
\caption{Real and imaginary parts of the $\bar{K}N$ scattering
amplitude in the isospin $I=0$ channel in the region of the
$\Lambda(1670)$ resonance.
\label{fig:KN0}}
\end{figure}

In Fig.~\ref{fig:KN0}, we show the real and imaginary parts of the $I=0$
scattering amplitude, normalized as in the Partial Wave Analysis
of Ref.~\cite{gopal77}. Remarkably, the amplitudes shown by 
the solid lines, which are obtained
using the low-energy parameters in Eq.~(\ref{eq:params}),
show the resonant structure of the $\Lambda(1670)$
appearing at about the right energy and with a similar size
compared to the experimental analysis\cite{gopal77}.
The position of the resonance is quite
sensitive to the parameter $a_{K\Xi}$ and moderately sensitive to 
$a_{\eta \Lambda}$. Hence, without spoiling the nice agreement at low
energies, which is not sensitive to $a_{K\Xi}$, we 
exploit the freedom in the parameters by choosing $a_{K\Xi}=-2.70$, 
moving the resonance closer to its experimental
position (dashed lines).

SU(3) symmetry, partly broken here due to the use
of physical masses, demands a singlet and an octet of resonances.
Within $S=-1$, we have already identified
the singlet $\Lambda(1405)$ and the $I=0$ member of
the octet, the $\Lambda(1670)$. 
Since we found the partial decay widths and couplings\cite{ORB01}
of the $\Lambda(1405)$ to ${\bar K}N$ states
and the $\Lambda(1670)$ to $K\Xi$ states to be very large, one is naturally 
led to identify these two
resonances as a ``quasibound" ${\bar K}N$ and $K\Xi$ state,
respectively.

Searching for the $I=1$ member of the octet, 
we find that the $I=1$ amplitudes in our model are smooth and show no
trace of resonant behavior, in line with experimental observation.
To explore this issue further we conducted a search for the poles of
the ${\bar K}N \to {\bar K}N$ amplitudes in the second Riemann sheet
and find two poles in the $I=0$ amplitude
($1426 + {\rm i} 16, ~ 1708 + {\rm i} 21 $), corresponding to the
$\Lambda(1405)$ and the $\Lambda(1670)$, and one in the $I=1$ amplitude 
$(1579 + {\rm i} 296)$,
corresponding - most likely - to the resonance $\Sigma(1620)$. The large
width found for this resonance may explain why we saw no trace 
of it in the scattering amplitudes.

\section{Strangeness $S=0$ sector}

\begin{figure}
\begin{minipage}{5.5cm}
\centerline{
\psfig{file=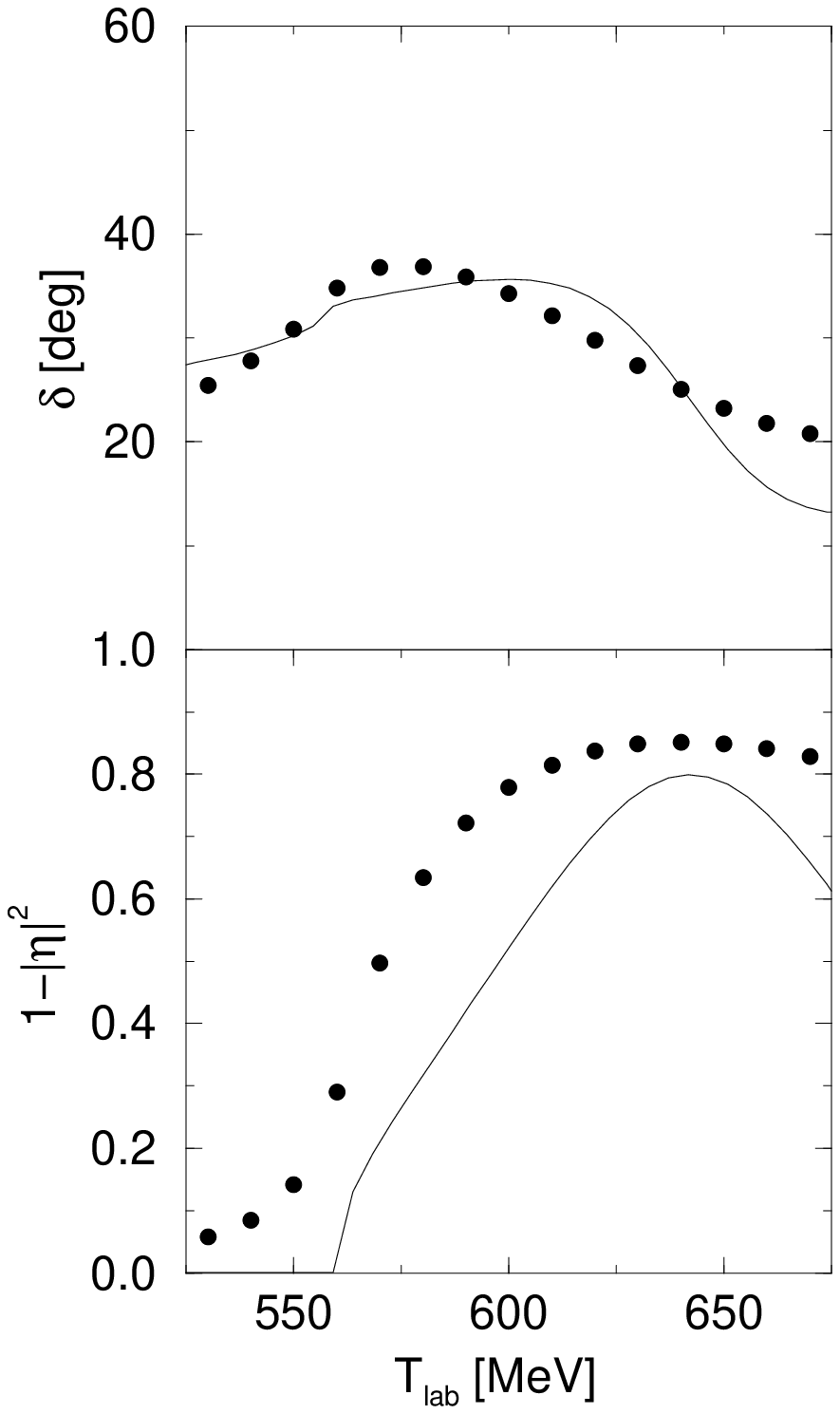,height=8cm,angle=0,silent=} 
}
\caption{Phase-shifts and inelasticities for $\pi N$ scattering in
the isospin $I=1/2$ channel. Taken from Ref.~\protect\cite{NPOR00}.
\label{fig:phase}}
\end{minipage}\hspace{1cm}
\begin{minipage}{5.5cm}
\centerline{
\psfig{file=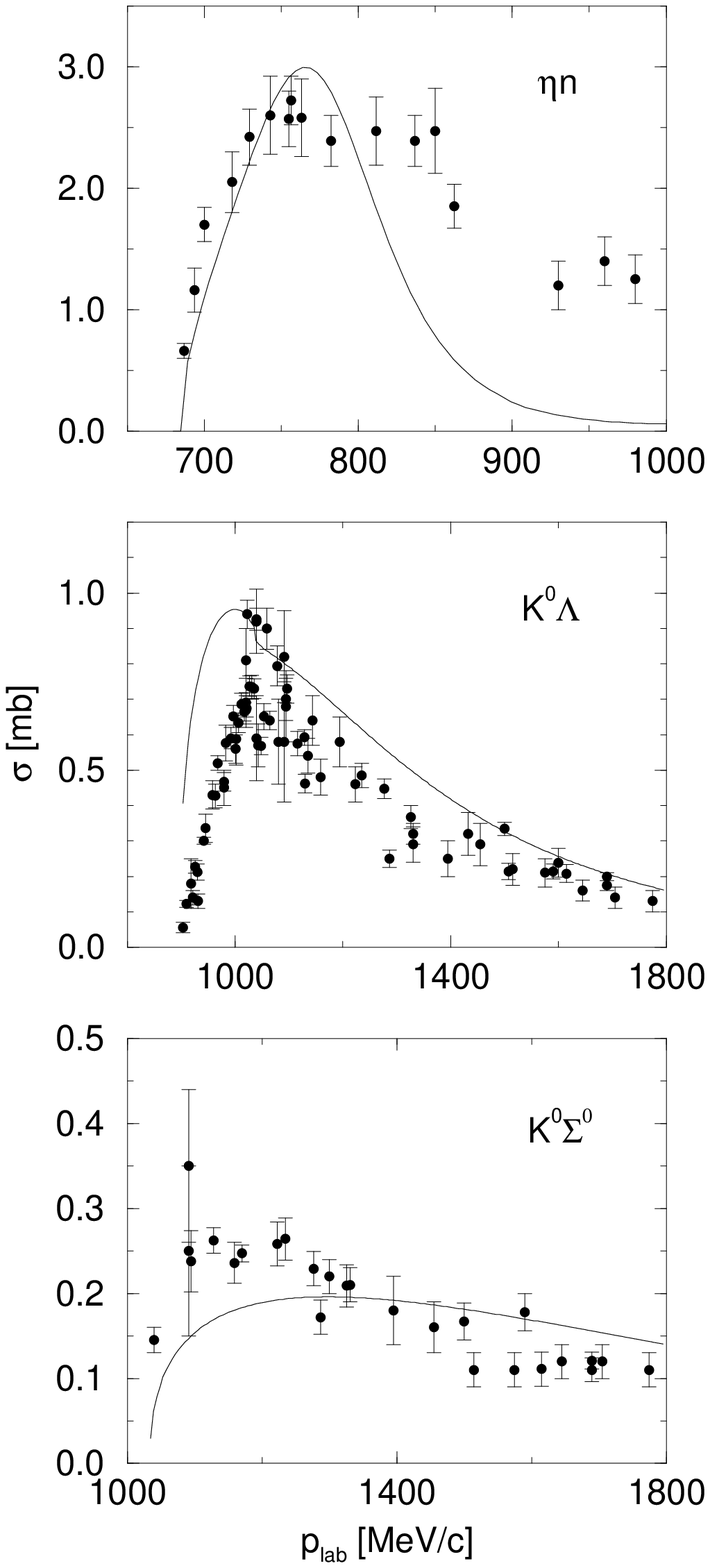,height=8cm,angle=0,silent=} 
}
\caption{Cross sections for the $\pi^- p \to \eta n$, $K^0
\Lambda$ and $K^0 \Sigma^0$ reactions as function of the $\pi^-$
laboratory momentum. Taken from Ref.~\protect\cite{NPOR00}.
\label{fig:cross}}
\end{minipage}
\end{figure}

The success in the s-wave $S=-1$ sector with the
lowest-order chiral Lagrangian and only one-parameter\cite{OR98} encouraged
us to study the $S=0$ sector in the vicinity of the $N(1535)$ resonance. In
this case, however, it was not possible to reproduce the elastic $\pi N$
scattering observables and a few inelastic cross sections with only one
parameter.
In ref.\cite{NPOR00}, we adopted a method similar to the dimensional
regularization scheme, by using a fixed physical cut-off of 1
GeV, large enough to study the energy region of the $N(1535)$ resonance, and
adding a subtraction constant $a_l$ to the loop function $G_l$.
The values of the decay constants were taken different for
$\pi N$ channels ($f_\pi=93$ MeV), $K Y$ channels ($f_K=1.22 f_\pi$) and
$\eta N$ channels ($f_\eta=1.3 f_\pi$). Our $S=0$ model
has then four parameters, $a_{\pi N}$, $a_{\eta N}$, $a_{K
\Lambda}$ and $a_{K \Sigma}$, which are fitted to the phase-shifts and
inelasticities for $\pi N$ scattering in isospin $I=1/2$ around the energy
region of the $N(1535)$ resonance, as shown in Fig.~\ref{fig:phase}, and to
the inelastic cross section data, as shown in Fig.~\ref{fig:cross}. 
The analysis performed in Ref.~\cite{NPOR00} leads to a dynamical generation
of the $N(1535)$ state with a total decay width of $\Gamma
\simeq 110$ MeV, divided between $\Gamma_\pi\simeq 43$ MeV and $\Gamma_\eta\simeq 67$ MeV,
compatible with present data within errors.
This simplified model allows one to tackle
other issues, such as the evaluation
of the $\pi^0 N^* N^*$ and $\eta N^* N^*$ couplings, which elucidates
the question of whether the positive parity $N$ and the
negative parity $N^*$ are chiral partners when chiral symmetry is
restored\cite{NPOR00}.

We note that a recent work, which also uses the chiral Lagrangian
together with a Bethe-Salpeter unitarization scheme, achieves a 
satisfactory description of the $S=0$ s-wave observables in the $I=1/2$
sector from $\pi N$ threshold up to 
2 GeV (similar in quality to our $S=-1$ description) 
by fitting 12 parameters\cite{juan01}, generating the
both the $N(1535)$ and $N(1650)$ $S_{11}$ resonances in the process.

\section{Strangeness $S=-2$ sector}

Within SU(3) the chiral meson-baryon Lagrangian naturally extends to the
$S=-2$ sector.  Thus, the dynamics of the ${\bar K} \Lambda$ or $\pi \Xi$
interaction can be predicted within the same approach. Due to the experimental 
difficulties there are very few scattering data on such reactions. Furthermore,
little is known about $\Xi$ resonances since they can only be produced 
as part of a multiparticle final state with small production cross sections.
Nevertheless, given the success of the approach presented 
here in generating most of the $(70,1_1^-)$ octet of $1/2^-$ excited baryon states
in the $S=0$ and $S=-1$ sector, it is reasonable to assume that its missing member
in the $S=-2$ realm, a $\Xi^*$ resonance, will be produced as well.
Among the observed states, two "suspects" stand out: the $\Xi$(1620) and the $\Xi$(1690), 
with unknown spin and parity. The 
$\Xi$(1690) might be the more plausible candidate since it has strong
couplings to the ${\bar K} \Lambda (\Sigma)$ channels which may serve 
as identifying features.  

\section{Conclusions}

We have extended the previously reported $S=-1$ meson-baryon scattering 
approach\cite{OR98} to higher energies of up to 2
GeV. This study has revealed the appearance of two other
resonances, the $\Lambda(1670)$ and the $\Sigma(1620)$, both belonging to
the octet, dynamically generated as $K\Xi$
states.

The  $S=0$ sector has been more difficult to describe
with only a few parameters, although the model presented here 
is able to reproduce the $\pi N$ data moderately well
around the energy of the $N(1535)$ and allows one to tackle other
issues related to the $N(1535)$, such as its couplings
to $\pi$ and $\eta$ mesons\cite{NPOR00}. 
A complete description from low to high
energies, as presented here for the $S=-1$ sector, has not been possible within
our simple approach, but it has been achieved in the $I=1/2$ sector
using an extended parameter set\cite{juan01}.
Given the success of this method of producing the members of the $1/2^-$ octet
of excited states in the $S=0$ and $S=-1$ sectors,
it is reasonable to assume that an extension to the $S=-2$ realm will 
generate the missing octet member, a $\Xi$ resonance in the energy region
of 1600-1700 MeV. Confirmation of such a state would further
demonstrate the power of combining
the chiral meson-baryon Lagrangian with non-perturbative unitarization techniques.




\section*{Acknowledgments}

This work has been supported by DGICYT contract
numbers BFM2000-1326, PB98-1247, by the Generalitat de Catalunya project
SGR2000-24, by the EU TMR network Eurodaphne, contract no.
ERBFMRX-CT98-0169, and by US-DOE grant DE-FG02-95ER-40907.


\begin{thebibliography}{99}
%
%
%
%


\bibitem{Ga85} J. Gasser and H. Leutwyler, Nucl. Phys. B250 (1985) 465

\bibitem{Pi95} A. Pich, Rep. Prog. Phys. 58 (1995) 563 
\bibitem{Eck95}G. Ecker, Prog. Part. Nucl. Phys. 35 (1995) 1 
\bibitem{Be95} V. Bernard, N.Kaiser, U.G. Meissner, Int. J. Mod. Phys. E4 (1995) 193

\bibitem{OOP99} J. A. Oller, E. Oset, and J. R. Pel\'aez, Phys.\ Rev.\
Lett.\ {80} (1998) 3452; Phys.\ Rev.\ D {59} (1999) 074001;
 erratum Phys.\ Rev.\ D { 60} (1999) 099906

\bibitem{nsd} J.A. Oller and E. Oset, Phys. Rev. D60 (1999)
074023 

\bibitem{Kai95} N. Kaiser, P. B. Siegel and W. Weise, Nucl. Phys. A594
(1995) 325; N. Kaiser, T. Waas and W. Weise, Nucl. Phys. A612 (1997) 297

\bibitem{OR98} E. Oset and A. Ramos, Nucl. Phys. A635 (1998) 99

\bibitem{OM01} J.A. Oller and U.G. Meissner, Phys. Lett. B500 (2001) 263

\bibitem{juan01} J. Nieves and E. Ruiz Arriola, hep-ph/0104307

\bibitem{report} J.A. Oller, E. Oset, A. Ramos,
Prog. Part. Nucl. Phys. 45 (2000) 157


\bibitem{spec} R.J. Hemingway, Nucl. Phys. B253
(1985) 742

\bibitem{gopal77} G.P. Gopal et al., Nucl. Phys. B119 (1977) 362


\bibitem{ORB01} E. Oset, A. Ramos and C. Bennhold, in preparation

\bibitem{NPOR00} J.C. Nacher, A. Parre\~no, E. Oset, A. Ramos, A. Hosaka
and M. Oka, Nucl. Phys. A678 (2000) 187

\end{thebibliography}
\end{document}